\documentclass[12pt, draftclsnofoot, onecolumn]{IEEEtran}
\usepackage[usenames,dvipsnames,svgnames,table]{xcolor}
\usepackage{times}
\usepackage{epsfig}
\usepackage{graphicx}
\usepackage{amsmath}
\usepackage{bbm}
\usepackage{amssymb}
\usepackage{amsxtra}
\usepackage{epstopdf}
\usepackage{rawfonts}
\usepackage{times}
\usepackage{url}
\usepackage{here}
\usepackage{cite}
\usepackage{caption}
\usepackage{bbm}
\usepackage{array}
\usepackage{longtable}
\usepackage{array,booktabs,enumitem}%
\usepackage{etoolbox}

\newcolumntype{P}[1]{>{\endgraf\vspace*{-\baselineskip}}p{#1}}

\usepackage{booktabs}

\let\mybibitem\bibitem
\renewcommand{\bibitem}[1]{%
  \ifstrequal{#1}{nature}
    {\color{blue}\mybibitem{#1}}
    {\color{black}\mybibitem{#1}}%
}
\begin{document}
%
\title{A Multi-Game Framework for Harmonized LTE-U and WiFi Coexistence over Unlicensed Bands}

\author{\authorblockN{ Kenza Hamidouche$^{1,2}$, Walid Saad$^2$, and  M\'erouane Debbah$^{1,3}$}\\ \authorblockA{\small
$^{1}$ CentraleSup\'elec, Universit\'e Paris-Saclay, France, {\tt\footnotesize kenza.hamidouche@centralesupelec.fr}\\
$^2$ Wireless@VT, Bradley Department of Electrical and Computer Engineering, Virginia Tech, USA, {\tt\footnotesize walids@vt.edu}\\
$^{3}$ Mathematical and Algorithmic Sciences Lab, Huawei France R\&D, France, {\tt\footnotesize merouane.debbah@huawei.com} \vspace{-0.6cm}}%
   \thanks{This research was supported by the ERC Starting Grant 305123 MORE (Advanced Mathematical Tools for Complex Network Engineering), the ANR project:  WisePhy: S\'ecurit\'e pour les communications sans fil \`a la couche physique, the U.S. National Science Foundation under Grants AST-1506297, CNS-1513697, and CNS-1460316.}
 }
\date{}


%


\maketitle



\begin{abstract}
The introduction of LTE over unlicensed bands (LTE-U) will enable LTE base stations (BSs) to boost their capacity and offload their traffic by exploiting the underused unlicensed bands. However, to reap the benefits of LTE-U, it is necessary to address various new challenges associated with LTE-U and WiFi coexistence. In particular, new resource management techniques must be developed to optimize the usage of the network resources while handling the interdependence between WiFi and LTE users and ensuring that WiFi users are not jeopardized. To this end, in this paper, a new game theoretic tool, dubbed as \emph{multi-game} framework is proposed as a promising approach for modeling resource allocation problems in LTE-U. In such a framework, multiple, co-existing and coupled games across heterogeneous channels can be formulated to capture the specific characteristics of LTE-U. Such games can be of different properties and types but their outcomes are largely interdependent. After introducing the basics of the multi-game framework, two classes of algorithms are outlined to achieve the new solution concepts of multi-games. Simulation results are then conducted to show how such a multi-game can effectively capture the specific properties of LTE-U and make of it a ``friendly'' neighbor to WiFi.
\end{abstract}

\IEEEpeerreviewmaketitle

\vspace{-0.3cm}

\section{Introduction}
Mobile data traffic has experienced an exponential growth during the last decade and it is expected to continue increasing by 1000 times in the upcoming five years \cite{cisco}. Such multimedia traffic demand with a high quality-of-service (QoS) requirement has strained the capacity of current wireless networks and gave rise to new challenges such as backhaul management, resource allocation, and interference management. 

One promising approach to cope with the continuously increasing data traffic and meet the stringent QoS of emerging wireless services, is by enabling LTE base stations (BSs) to exploit the readily available unlicensed spectrum\cite{qualcomm}. Although offloading part of the LTE traffic to the unlicensed bands can considerably increase the performance of cellular networks, the heterogeneity of the devices coupled with the disparity of the medium access protocols that are used by the WiFi and LTE-U access points (APs), rise new challenges that need to be addressed before these networks can reach their true potential \cite{zhang2015lte}. In fact, if the LTE BSs transmit continuously over the unlicensed bands, the WAPs will experience high interference from the BSs resulting in large backoff durations that prevent the WAPs from serving the regular WiFi users. 

Thus, to avoid the deterioration of the WiFi network's performance, cellular BSs must adapt their offloaded traffic based on the WAPs activities. This will in turn impact the QoS guarantees that can be provided by the BSs as it will depend on the varying WiFi traffic load on the unlicensed bands. Hence, new resource management techniques are necessary to ensure a harmonious coexistence of the two technologies. To this end, the traditional resources allocation mechanisms at both LTE and WiFi networks need to be modified to account for the interdependence between the devices in the two networks. The coupled resource management decisions of the users at the inter-network and intra-network level, as well as the heterogeneity of the LTE-U system motivate the adoption of game-theoretic solutions\cite{han2012game}.  

Game theory is a set of mathematical tools used to model and analyze the complex interactions between independent rational players with coupled objectives. It has been extensively used in communications, networking and signal processing, and has proven to be useful for addressing many problems in these systems \cite{han2012game}. In particular, game theory has been recently used to design efficient algorithmic solutions for resource allocation problems in LTE-U \cite{etkin2007spectrum,teng2014sharing,chen2016echo,bennis2013cellular}. 

In \cite{etkin2007spectrum}, the authors modeled the sharing problem of unlicensed bands among operators as one-shot and repeated noncooperative game. In \cite{teng2014sharing}, a similar problem is also formulated as a repeated noncooperative game for more general utility functions. The work in \cite{teng2014sharing} shows that operators can reach a subgame perfect Nash equilibrium. The authors in \cite{chen2016echo,Opt16Chen} formulated the unlicensed spectrum allocation problem with uplink-downlink decoupling as a noncooperative game in which the SBSs are the players that select the unlicensed channels over which they serve their users. The goal of the SBSs is to optimize the uplink and downlink sum-rate while balancing the licensed and unlicensed spectrums between the users. A learning algorithm is proposed in \cite{bennis2013cellular} for the joint interference management and traffic offloading problem at the unlicensed bands. The authors in \cite{liu2014small} use classical optimization tools to define the optimal unlicensed spectrum allocation to the BSs. 

Despite being interesting, most of these works consider models in which the dependence between the LTE and WiFi networks is ignored. In fact, this existing body of work \cite{etkin2007spectrum,teng2014sharing,chen2016echo,bennis2013cellular,liu2014small,Opt16Chen} has treated allocation problems on each network independently. In fact, in LTE-U, the inter-network interactions can significantly impact the outcome of the resource allocation models. For instance, a dual-mode BS that has access to both networks can benefit from the available information on one network to improve its performance on the second network. This is a critical point especially for BSs as they must ensure the expected QoS by their users, while over unlicensed bands, there are no guarantees regarding the delivery. Moreover, the BSs must account for the payoff of the WAPs to reach a fair coexistence on the unlicensed bands and adapt the allocation of the licensed bands accordingly.

The main contribution of this paper is to provide a new game-theoretic framework that allows to address and capture the following characteristics of LTE-U: 1) disparity in the used access protocols by the LTE BSs and the WAPs on the unlicensed bands, 2) heterogeneity of the network devices with the integration of dual-mode BSs and the existing WAPs and BSs, 3) different access bands and QoS requirements, and 4) need of WiFi-aware algorithms to ensure a fair coexistence on the unlicensed spectrum. To this end, we introduce a novel multi-game framework, that generalizes existing game-theoretic solutions and enables one to capture all the specific characteristics of LTE-U. In this regard, we first provide an overview of the main challenges of LTE-U. Then, we describe the proposed multi-game model and new solution concepts that are desirable in such networks. Finally, we provide simulations results to validate the proposed multi-game framework and we discuss future opportunities in this space.

\section{LTE over Unlicensed: Overview and Challenges}
\label{desc}
\begin{figure}[H]
\centering
\includegraphics[scale=0.4]{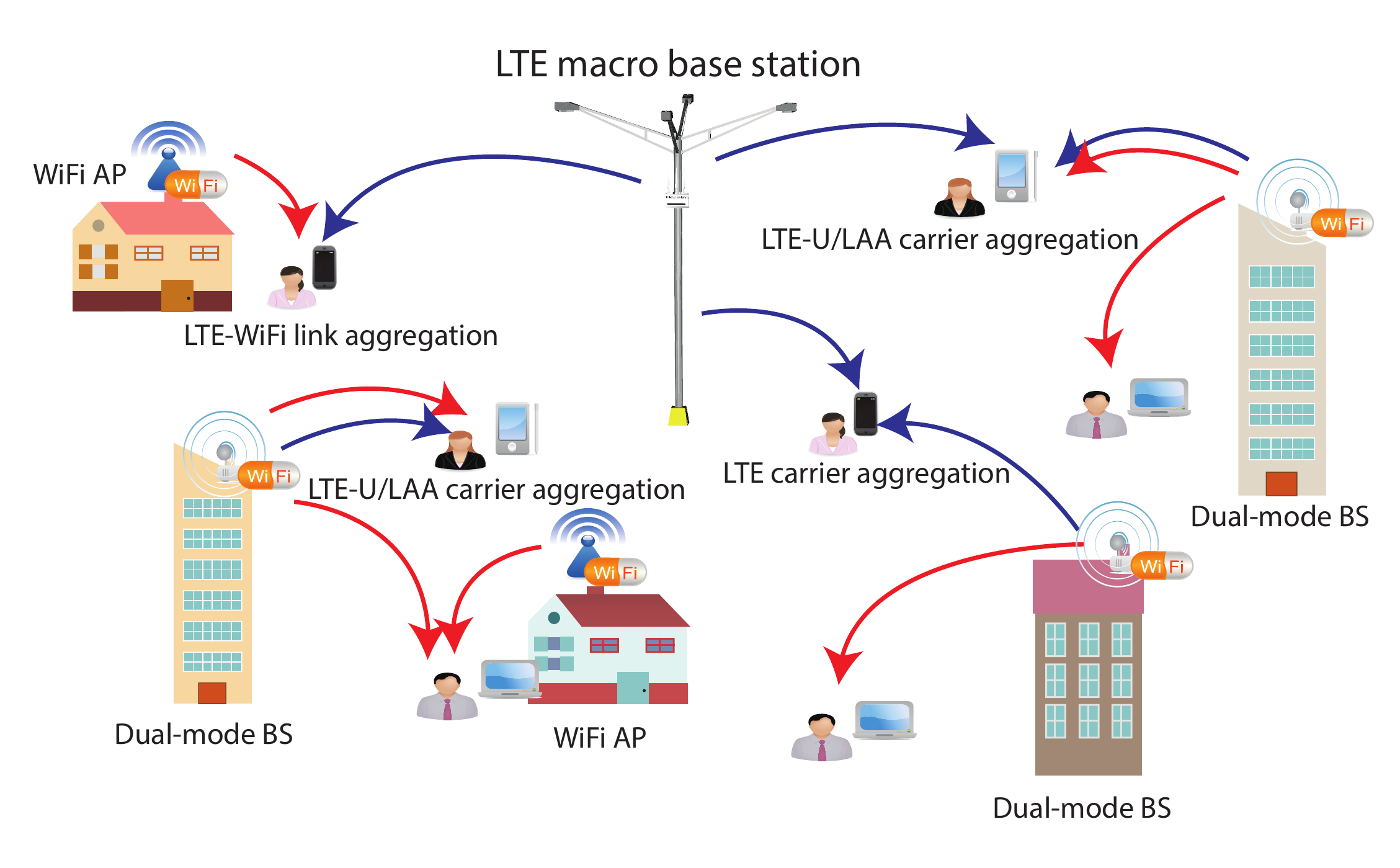}\vspace{-0.1cm}
\caption{LTE-U network model.}\vspace{-0.3cm}
\label{fig:model}
\end{figure}

LTE-U allows an LTE system to exploit the unlicensed bands so as to offload a portion of its traffic, thus reducing the overall congestion over the licensed spectrum. This is ensured via the use of carrier aggregation technologies as well as the development of dual-mode BSs that can transmit simultaneously over licensed and unlicensed bands. Multiple techniques have been advocated recently by academia and industry to reach a harmonious coexistence over the unlicensed bands by preventing the LTE BSs from harming the WiFi users. Such techniques can be classified into two categories illustrated in Figure \ref{fig:model},\cite{qualcomm}:
\begin{itemize}
\item \emph{LTE-WiFi link aggregation (LWA)}: LWA is enabled by integrated LTE/WiFi BSs equipped with a WiFi interface and an LTE interface. Thus, the BSs are able to transmit simultaneously over the licensed and unlicensed bands and then use link aggregation to form a larger transmission bandwidth. In LWA, the transmitting BSs use the same medium access technique as WiFi known as listen-before-talk (LBT), in which users sense the channels and transmit only if the medium is sensed to be idle. 
\item \emph{LTE-U and licensed assisted access (LAA)}: LAA uses LTE carrier aggregation technologies to combine LTE in the licensed bands with LTE in the unlicensed bands. This is ensured by dual-band BSs that can adaptively switch between licensed and unlicensed bands via the same LTE interface. In LAA, an LBT-like technique is used to access the unlicensed bands with longer sensing periods. Based on the observed medium activities, the BSs can then determine the fraction of time they use the channels. On the other hand, LTE-U is a more open model in which BSs do not have any regulation constraints regarding the fraction of time they use the unlicensed channels or the caused interference to the WiFi users.
\end{itemize}
 
LTE-U does not require any changes in the LTE air interface protocol, enabling an early deployment of the technology in many countries including the USA, South Korea, China, and India. However, regular users (i.e., the common people using WiFi), are starting to raise serious objections about cellular operators taking over their bands \cite{google}. In fact, without regulatory requirements in LTE-U, LTE BSs can significantly impact the performance of WiFi users as WiFi users cannot access the  channels when they sense a high interference level from the continuously transmitting BSs. Thus, LTE-U introduces new challenges that must be addressed to make its impact on WiFi of equal magnitude as that of another WiFi network:

\begin{itemize}
\item \emph{Heterogeneity in device types and access protocols:} The heterogeneity of the APs in cellular networks appears in the cell coverage and the transmit power which are typically lower at the WAPs compared to the LTE BSs. Moreover, the APs use different medium access control methods. To access the channels, WAPs use carrier sense multiple access with collision avoidance (CSMA/CA), in which WAPs  sense the channel and attempt to avoid collisions. When the sensed interference level exceeds a predefined threshold or a collision occurs, the transmitting WAPs enter a period of binary exponential backoff prior to attempting another transmission. On the other hand, BSs scheduling techniques, such as orthogonal frequency division multiple access (OFDMA) that can guarantee QoS by dividing the bandwidth into a number of physical resource blocks allocated to multiple BSs at the same time.  
\item\emph{Traffic offloading and QoS requirements:} In contrast to LTE networks in which the resources are preallocated to maximize the QoS of the users, WiFi networks do not provide any QoS guarantees. Consequently, when LBT is used by the BSs, a random offload of the content by the BSs may result in a discontinuous service and a degraded QoS for the LTE users. Thus, new resource allocation approaches must be developed in the LTE network to allocate the network resources while being cognizant of the type and the requirements of the contents in terms of data rate as well as the load of the WiFi network.
\item \emph{Cross-technology interference and collisions:} When LTE and WiFi exist over the unlicensed spectrum, the WAPs are subject to excessive interference if the power regulation in WiFi is not respected by the LTE users or the number of coexisting BSs is large. Thus, the allocation of unlicensed channels to the BSs must consider the impact of the offloaded traffic on the throughput of the WAPs. In fact, when the interference is sensed higher than a predefined threshold, the WiFi users are not able to decode the transmitted content correctly and thus stop the transmission and backoff for an exponentially increasing time duration. This results in a deterioration of the WiFi performance if the LTE users keep transmitting over the unlicensed channels without taking into account the incumbent users of the WiFi network.
\end{itemize}
Given the characteristics of LTE and WiFi networks, the resource management decisions of the APs are highly interdependent. For example, this can appear through the data rate achievable by the APs which depends on all the APs transmitting over the same channel. This interdependence between the resource management decisions of the APs makes game theory a natural tool to study and analyze such networks. In this regard, in \cite{etkin2007spectrum,teng2014sharing,chen2016echo,bennis2013cellular}, game-theoretic solutions were proposed to address the new challenges of LTE-U for a fair coexistence of LTE and WiFi networks.   

However, in traditional game-theoretic analysis of LTE-U \cite{etkin2007spectrum,teng2014sharing,chen2016echo,bennis2013cellular}, the resource allocation problems are formulated as single games. Hence, one typically analyzes the resource allocation problems at the WiFi and LTE networks \emph{independently}, ignoring the impact of the interdependence of the two networks. In other words, the outcome of the resource allocation and the possible available information on one of the networks is not exploited at the other network. Such information can be easily made available by the dual-mode APs that belong to both WiFi and LTE networks and thus, can be used as a feedback at one of the networks to improve its performance. 

Moreover, in contrast to similar ideas such as cross-system learning in \cite{bennis2013cellular}, the allocation of resources in LTE-U must account for the multiple coexisting LTE and WiFi systems as well as their impact on one another. This is necessary to capture the heterogeneity of the APs, their different utilities and objectives in the networks depending on the resources for which the players contend such as licensed and unlicensed bands. These specific characteristics of LTE-U that are ignored in \cite{bennis2013cellular}, can enable the BSs to adapt their resource allocation methods by exploiting the feedback from the other games (of possibly different type), and thus, not to jeopardize the data rate of the incumbent WAPs in the unlicensed bands. 

To capture and leverage the multi-mode transmission that is specific to LTE-WiFi networks, new approaches should be developed and used to address the new challenges in these networks. In the next section, we introduce a holistic game-theoretic framework to model the resource management problems while fully exploiting the properties that are specific to LTE-WiFi networks. 
\section{Multi- Games for LTE-U}
\label{game}
Here, we propose the framework of \emph{multi-game theory} as a new tool suitable to capture the inter-network and intra-network interactions between the users in LTE-WiFi networks. A multi-game $\mathcal{G}$ is a game that is composed of multiple interdependent games $\mathcal{G}=\{\mathcal{G}_1,...,\mathcal{G}_N\}$. Each game $\mathcal{G}_1$ models a specific resource management problem at the WiFi level, LTE level or coupled level. 
Multi-games can be defined by three parameters, the players, the actions, and the utility function. The players can be BSs, WAPs, LTE user equipment (UEs), or WUEs, each of which is associated a set of actions from which it can choose it strategy. The goal of each player is to select the strategy that maximizes a utility function  that corresponds to its goal in the network. For example, BSs aim to maximize the experienced quality by their users over both licensed and unlicensed bands. On the other hand, based on the best-effort service, WiFi users would prefer to deliver a maximum amount of data. The utility is given by a function $u_{m}(a_{-m}, \{\boldsymbol{o}_i\}_{n=1,..,N \setminus n})$, which depends on the actions that are selected by all its co-players in the game $\mathcal{G}_n$ as well as the outcome of all the other games denoted $\boldsymbol{o}_i$ for each game $\mathcal{G}_i$. For instance in LTE-U, the throughput of a WiFi user is limited by the interference from both BSs and WiFi users transmitting over the same channel.

The defined multi-game framework for LTE-U have two key properties: 
\begin{itemize}
\item \emph{Heterogeneity of the games type:} Each of the formulated games to model one of the resource allocation problems in both networks can be of different type. For instance, the allocation of the unlicensed bands to the WAPs can be formulated as a cooperative game while the allocation of the licensed bands to the BSs as a two-sided matching game. When considering the power allocation problem of the APs, a noncooperative game can be more suitable. The problems can even be formulated as classical non-game theoretic optimization problems.
\item \emph{Interdependence between games:} In addition to the utilities of the players that depend on the outcome of the other games, the dependence between the games can appear in two other ways. Each game of the multi-game can be an inter-network or intra-network game, thus, a game can be played excursively by WiFi users, LTE users, or network operators but it can also be played by a combination of these players. Moreover, the set of actions of the players can be the same for many games such as the APs that can be selected by either WUEs, LTE UEs and UEs having both functionalities. 
\end{itemize}

Hence, in contrast to the popular two-level Stackelberg games in which the dependence between the games and the players appears only through the utility that is a function of the outcome of the other game, in the proposed multi-game model, in addition to the multi-level structure of the game, the dependence between the games can appear in various ways. 

In such multi-level games, the incumbent players at the considered network should have priority in accessing the network resources. Thus, when allocating unlicensed bands, the WiFi game is considered as the first-level game giving priority to the WAPs. Then, the players of each game respond to the decisions of the players at the higher-level games while simultaneously, anticipating the decisions of the players at the lower-level games and their possible reactions. The proposed multi-game framework  aligned with the idea of multi-resolution game theory in \cite{zhu2015game}. However, in contrast to the concepts in \cite{zhu2015game} which are mainly aligned with problems in security and resilience of control systems, we consider games of different types while focusing on LTE-U resource management.

\subsection{Multi-Game Solution Concepts}
Having presented the general multi-game model, the goal now is to define suitable solution concepts that can achieve an optimal resource allocation across the LTE and WiFi networks as well as the unlicensed bands. Nash equilibrium (NE) and stability are the two solution concepts that can be desirable depending on the formulated games. The NE is suitable for games such as noncooperative games while stability is defined in games such as cooperative and matching games in which groups of players are formed/matched. An \emph{equilibrium} represents the state in which none of the players can improve its utility by a given deviation, i.e., by selecting another action given that the actions of the other players are fixed. On the other hand, \emph{stability} refers to the state in which a player would not have the incentive to leave its group of players given that the other players have selected their partners or group mates. 

Applying classical equilibrium and stability notions to LTE-U will not be able to capture the interdependence between the games. Such conventional solutions do not account for the WiFi users' performance which is affected by the LTE users through the increase of collisions number. Thus, novel LTE-U oriented equilibrium and stability concepts must be defined to protect the WiFi users while maximizing the benefit of the LTE users on the unlicensed bands. To this end, we propose two solution concepts, \emph{multi-game equilibrium (MGE) and multi-game stability (MGS)} that can be characterized depending on the class of games used to formulate a given LTE-U problem. MGE and MGS represent the states in which all the WiFi users have reached a minimum target utility and all the LTE users have maximized their utility. In this state:
\begin{enumerate}
\item None of the WiFi users can reach the target utility by changing its strategy, given the actions of all LTE and other WiFi users fixed, 
\item None of the LTE users can improve its utility by changing its strategy, given the actions of all WiFi and other LTE users fixed.  
\end{enumerate}

At the equilibrium, all the WiFi users are satisfied by their utility and none of them can make any further improvements. Next, we present two new classes of algorithms to reach the equilibrium states in the defined multi-game.

\subsection{Algorithmic Solutions}
To allow the players to achieve the MGE and MGS, the classical algorithms that are used in single games cannot be applied anymore. In fact, such algorithms do not capture the interdependences between multiple heterogeneous games, and do not allow the players to exploit the specific properties in multi-games as some players can participate in multiple games at the same time and thus can exploit the feedback from one game to improve their performance on another game. Since the games are of multiple types, multiple algorithms must be combined to reach the equilibria or stable outcomes. More importantly, the algorithms developed must be distributed and provide adaptation to networks' changes. Following the introduced solution concept, we can define two classes of algorithms.
\subsubsection{Multi-game stability}
To reach the stable state in wireless networks when the resource allocation problems are formulated as matching games and coalitional formation games, decentralized algorithms such as deferred acceptance \cite{roth2008deferred} and merge-and-split \cite{saad2009coalitional} are used. These decentralized algorithmic solutions have been shown to significantly improve the performance of wireless networks with a low complexity. However, they are confined to single games. 

Hence, to reach the MGS outcome defined in the previous section, new cross-system algorithms must be developed. In such algorithms, dual-mode BSs must be able to observe their environment on the WiFi network and exploit this information to define their preferred actions on the LTE network which may correspond to the cells they want to join, or the other BSs with which they prefer to share the bands. Moreover, the multiplicity of the stable states in classical algorithms, introduces new algorithmic challenges that must be accounted for at each of the  algorithms that are defined to solve the games. In some cases, the players must be able to determine the worst possible outcome, i.e. stable state, of all the other algorithms and respond to it.
\subsubsection{Multi-game Nash equilibrium}
Nash equilibrium is the basic solution concept in noncooperative games. To reach this equilibrium, many learning algorithms can be used depending on the characteristics of the considered game \cite{young2004strategic}. Similar to the stability case, classical learning algorithms are designed for single games and do not account for the minimum utility requirements of the players and hence cannot be used to reach the introduced MGE. 

Consequently, new cross-system learning need to be developed for the multi-level games, to account for the interdependence between the games. In such algorithms, the players can explore the environment, learn their actions and exploit the feedback from the other bands and games. The players can then learn their strategies, estimate and update their utilities iteratively based on their observations. 

\subsection{Illustrative Multi-Game Models for LTE-U}
\label{ill}
\begin{figure*}[t]
	\centering
	\includegraphics[width=1\linewidth]{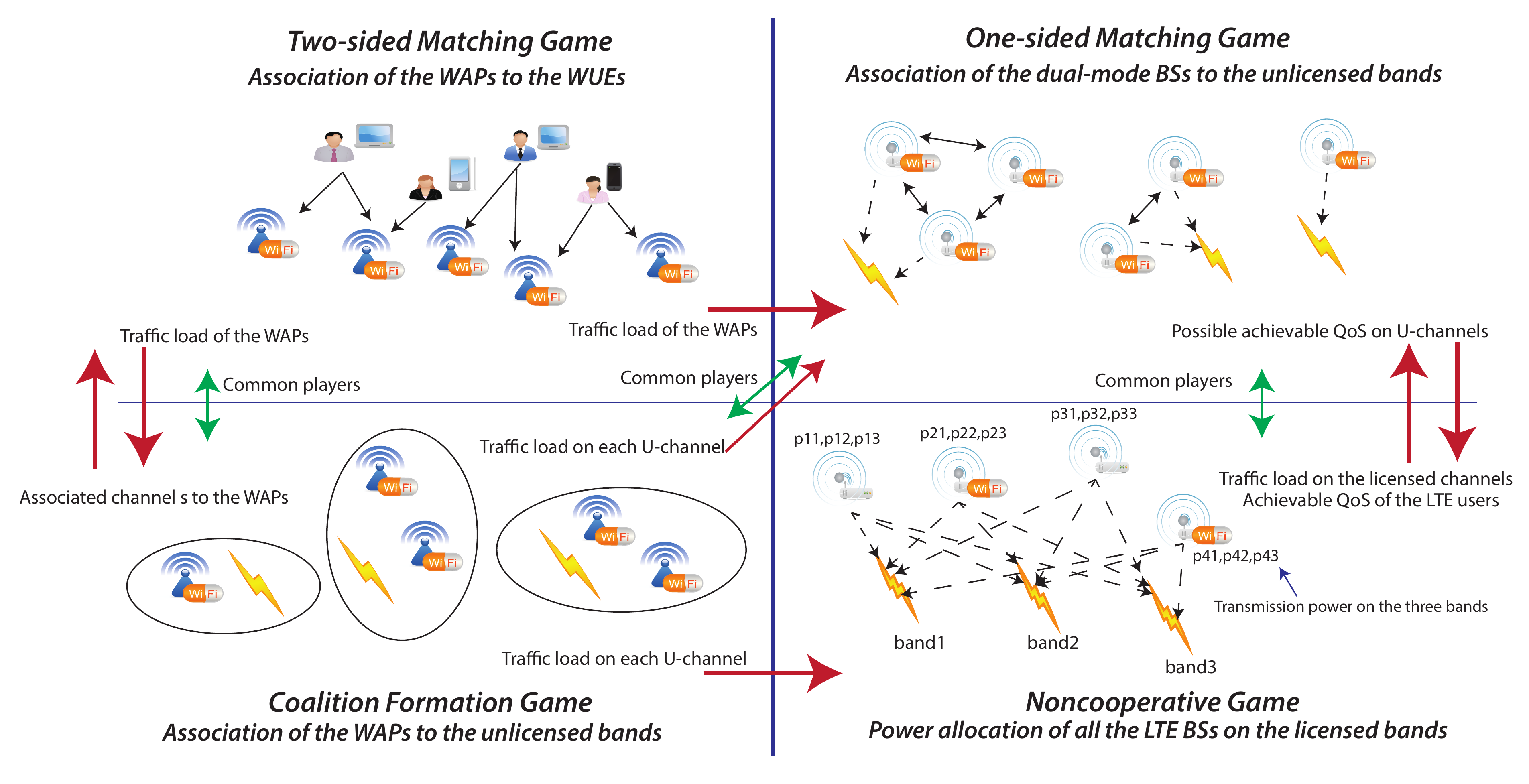}
	\caption{Illustrative multi-game model in LTE-U.}
	\label{fig:game}
\end{figure*}

Next, we show how the multi-game framework can be used to develop WiFi-aware resource allocation methods. An illustrative example is provided in Figure \ref{fig:game}, in which the multi-game framework is applied for the allocation of licensed and unlicensed bands to BSs and WAPs. This spectrum allocation problem depends on multiple parameters such as the traffic load generated by the BSs and the WAPs which in turn depends on the users associated to each BS and each WAP, respectively. In such network, the BSs, WAPs and users have different performance requirements and all aim to maximize their own utilities that are different. The interdependence between the four games is shown in Figure \ref{fig:game} using the red arrows that represent the information that affects the outcome of the game to which the arrow is directed. The green arrows connect every two games that have common players.

\subsubsection{Dynamic channel selection} The problem of unlicensed channel selection can be formulated as a multi-game composed of two game: 
\begin{itemize}
\item \emph{Association of the WAPs to the channels:} The players in this game are the WAPs and they aim to maximize a utility function which we assume to be the throughput which depends on the interference of the BSs and WAPs on each channel.
\item \emph{Association of the BSs to the channels:} In this game, the  players are the BSs and the utility function can be considered the achievable physical rate which depends also on the interference from the WAPs and BSs.
\end{itemize}
WAPs must be given priority as they are the incumbent users in the unlicensed band. The BSs then, account for the outcome of the WAPs game and respond to their strategies avoiding the deterioration of their throughput. For a more realistic model, we can account for the impact of the associated APs to the selfish users that aim to optimize their individual rate. Two additional games can be added in this case and are the WUEs-WAPs association and the LTE UEs-BSs association games, resulting in a four-levels multi-game. 

\subsubsection{Inter-operator spectrum sharing}
This game consists of a set of operators that own a set of BSs and contend for the unlicensed channels. Three games must be analyzed:
\begin{itemize}
\item \emph{Association of the WAPs to the unlicensed bands:} The utility of the WAPs in this game is the achievable throughput on each unlicensed channel.
\item \emph{Association of BSs to the licensed and unlicensed bands:} This game is played between the BSs and the bands. The strategies of the operators that own the BSs in this case, depend not only on the WAPs' selected actions but also on their subscribers, the price they pay, their minimum required QoS, as well as the load on the LTE network. In fact, if a user pays a low price and there is a high traffic on the LTE network, the operator would prefer to offload the user to the WiFi network to be able to guarantee the required QoS for the UEs that pay a higher price. This can be done only if the unlicensed spectrum is not saturated as the WAPs are the incumbent and have priority in using the spectrum. The utility of a BS is the achievable physical rate by all its users over a channel.
\item \emph{Association of the users to the operators:} This game that is played by the operators and user. Users aim to subscribe to the operator that offer the best deal in term of QoS/price which depends on the selected channels by the operators' BSs. On the other hand, operators prefer the users who pay more and for which the requested QoS can be satisfied. The utility of a user in this game is the difference between the achievable rate when served by a BS that deployed by a given operator and the cost for that service. On the other hand, the utility of an operator is the difference between the charged price to all the served users and the cost of serving them.
\end{itemize}

\section{Case Study: Multi-Games in LTE-U}
\label{sim}
To show the performance of the proposed multi-game framework, we perform simulations on a networks composed of dual-mode BSs and WAPs that contend for 10 unlicensed bands. Moreover, we consider a set of WUEs that aim to select the WiFi AP that maximize their physical rate. Thus, three coupled games are formulated as three different matching games. The network parameters are provided in Table I.
\begin{table}[H]
\scriptsize
\centering
\caption{\small  Network parameters}
\begin{tabular}{|l|l|}
\hline
{\bf Parameters}& {\bf Values}\\
\hline
Number of BSs & 40\\
\hline
Number of WAPs& 30\\
\hline
Mean time of a successful transmission&$5~~ \mu s$\\
\hline
Mean time of a collision& $1~~ \mu s$\\
\hline
Mean time of idle channel &$3~~\mu s$\\
\hline
Number of channels &10\\
\hline
RTS&20 bytes \\
\hline
CTS& 14 bytes\\
\hline
$DIFS$& $ 34~~ \mu s$\\
\hline
$SIFS$&$16 ~~\mu s$\\
\hline
Number of WUs&150\\
\hline
Transmit power of the WAPs&0.5W\\
\hline
Transmit power of the BSs&1W\\
\hline
\end{tabular}
\label{param}
\end{table}

\begin{itemize}
\item \emph{Association of WUEs to the WAPs:} This problem can be formulated as a two-sided matching game in which the WAPs aim to maximize the number of served UEs and the WUEs aim to maximize their physical rate.
\item \emph{Association of the WAPs to the unlicensed channels:} This problem  is formulated as one-sided matching in which the WAPs aim to autonomously form groups on the unlicensed channel that maximize their throughput.
\item \emph{Association of the LTE BSs to the unlicensed bands:} This problem is also formulated as a one-sided matching game in which the BSs select the channel that maximize the mean physical rate.
\end{itemize}
We can note from this formulation that WAPs participate in two games as players. Moreover, the utility of the players in each game depends on the outcome of the other games. For instance, the achievable physical rate by the WAPs depends on the unlicensed channels over which they serve their users in the second game and also on the associated BSs to that channel via the induced interference. 
\subsection{Multi-game vs. classical single game }
 \begin{figure}[H]
\centering
\includegraphics[scale=0.41]{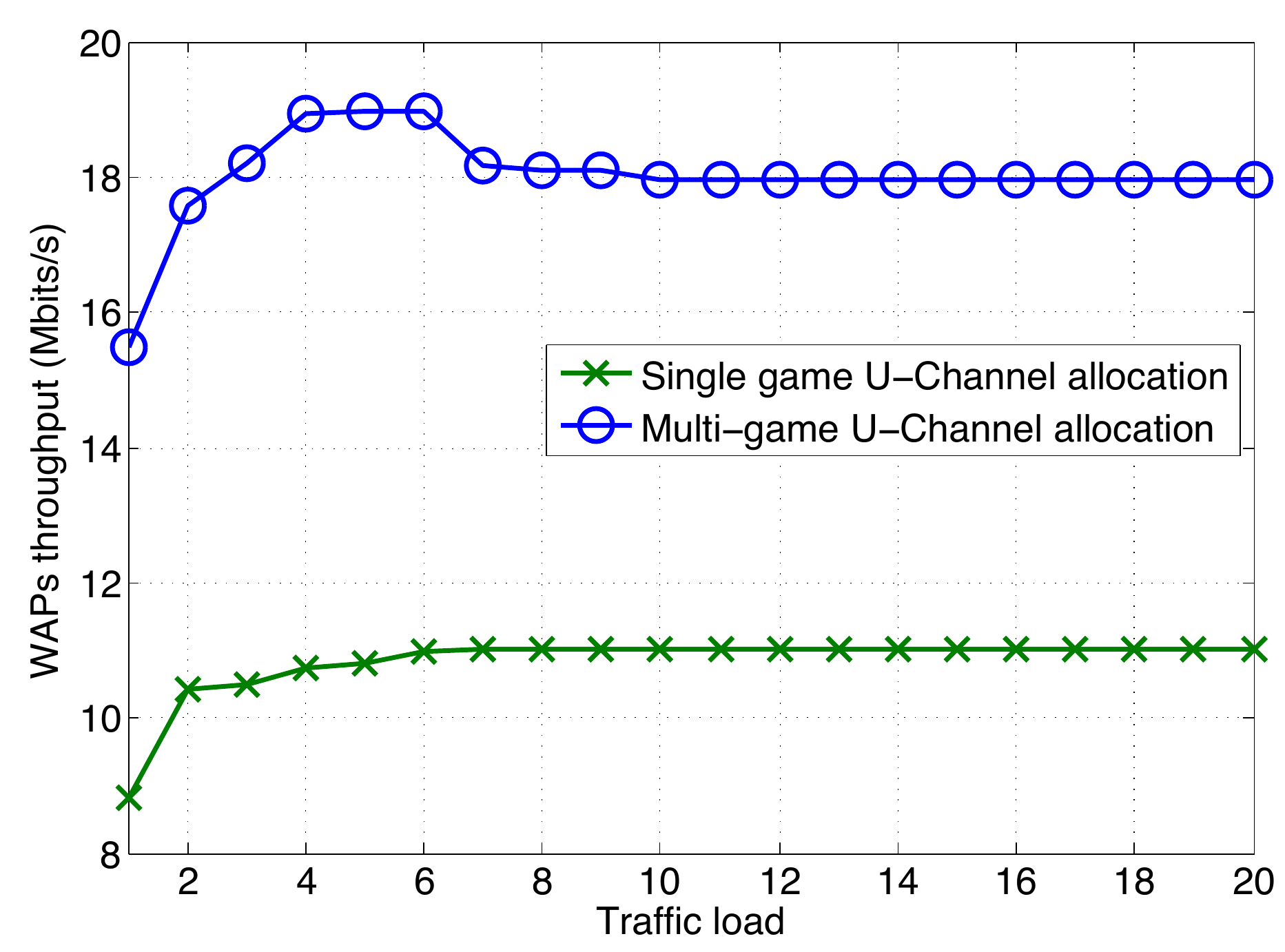}\vspace{-0.1cm}
\caption{Multi-game vs. single game in U-bands allocation.}\vspace{-0.4cm}
\label{res1}
\end{figure}
Figure \ref{res1} shows the sum-throughput of all the WAPs as a function of the traffic in the network. Two approaches are compared, the single game and multi-game models. In the former, there is no priority and all the BSs and WiFi can access the network similarly. In the latter, the WAPs are given the priority by playing the two first games and choosing the actions that maximize their utilities. Then, the BSs respond to their actions by using the unlicensed channels only when they are not used by the WAPs. From Figure \ref{res1}, we can see that the network saturates faster in the single-game model compared to the multi-game. This is due to the fact that the BSs can monopolize the network and transmit for an infinite time when the WAPs are not given priority. This results in a sum-throughput that is 50 \% lower  for the WAPs, in the single game compared to the multi-game case. 

\subsection{BSs utility}
\begin{figure}[H]
\centering
\includegraphics[scale=0.42]{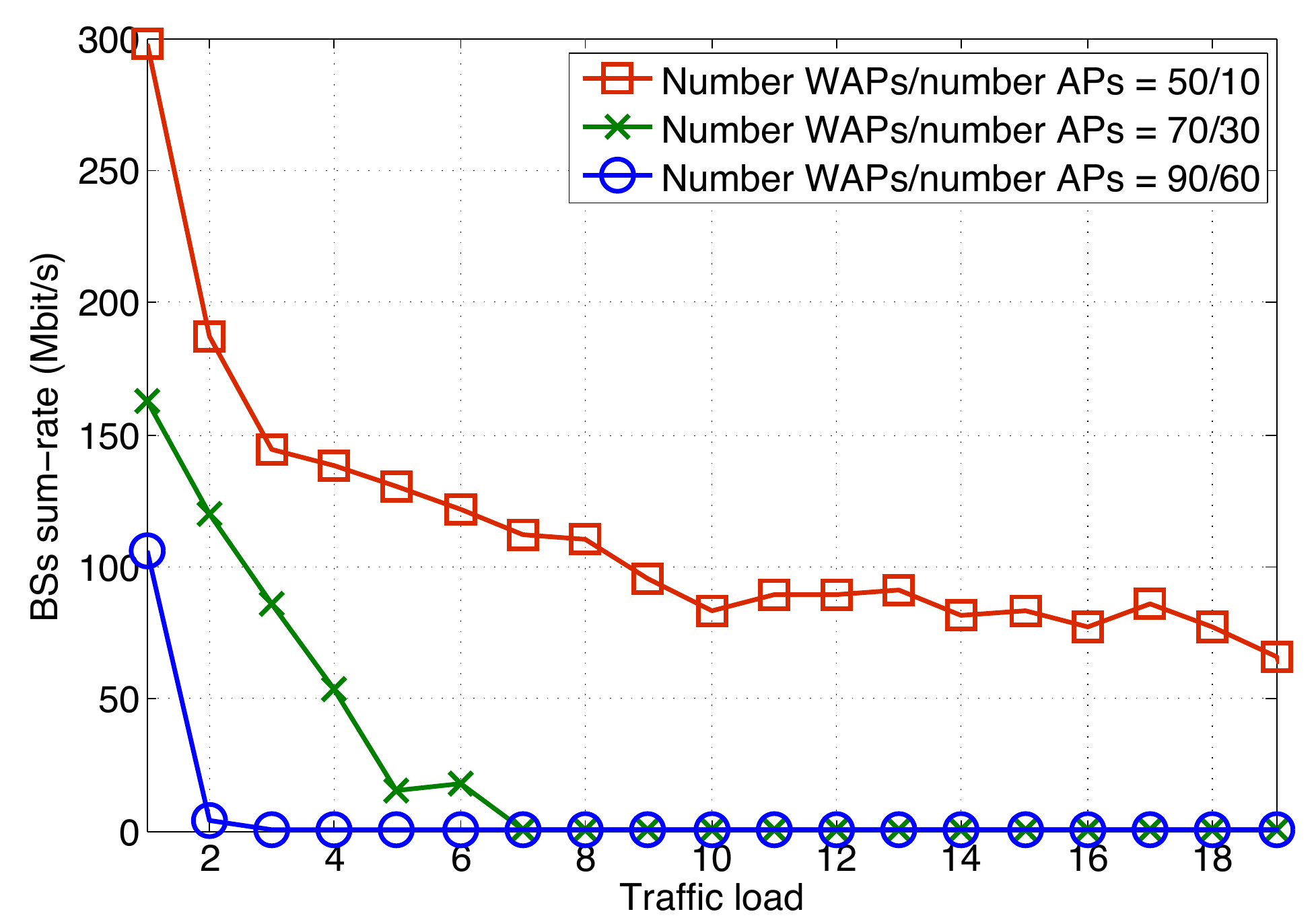}\vspace{-0.1cm}
\caption{BSs sum-rate with respect to the WiFi traffic load and number of WAP/APs.}
\label{res2}
\end{figure}
Figure \ref{res2} shows how the BSs adapt the amount of offloaded traffic to the unlicensed channels based on the number of WAPs that are considered in the network. In fact, the ability of adaptation to network changes at the BSs allows us to show how the proposed multi-game framework impacts the performance of the LTE-U network as compared to LBT and single game models. In particular, we show the sum-rate of the BSs when increasing the traffic load of the WAPs and the number of WAPs is the network. Figure \ref{res2} shows that as the number of WAPs increases, the sum-rate of the BSs decreases. This is due to the limited capacity of the unlicensed bands that saturates when increasing the load. Thus, based on the priority model, the BSs decrease the amount of offloaded traffic over the unlicensed bands as the WiFi traffic increases. 

\subsection{Multi-game vs. LBT}
 \begin{figure}[H]
\centering
\includegraphics[scale=0.42]{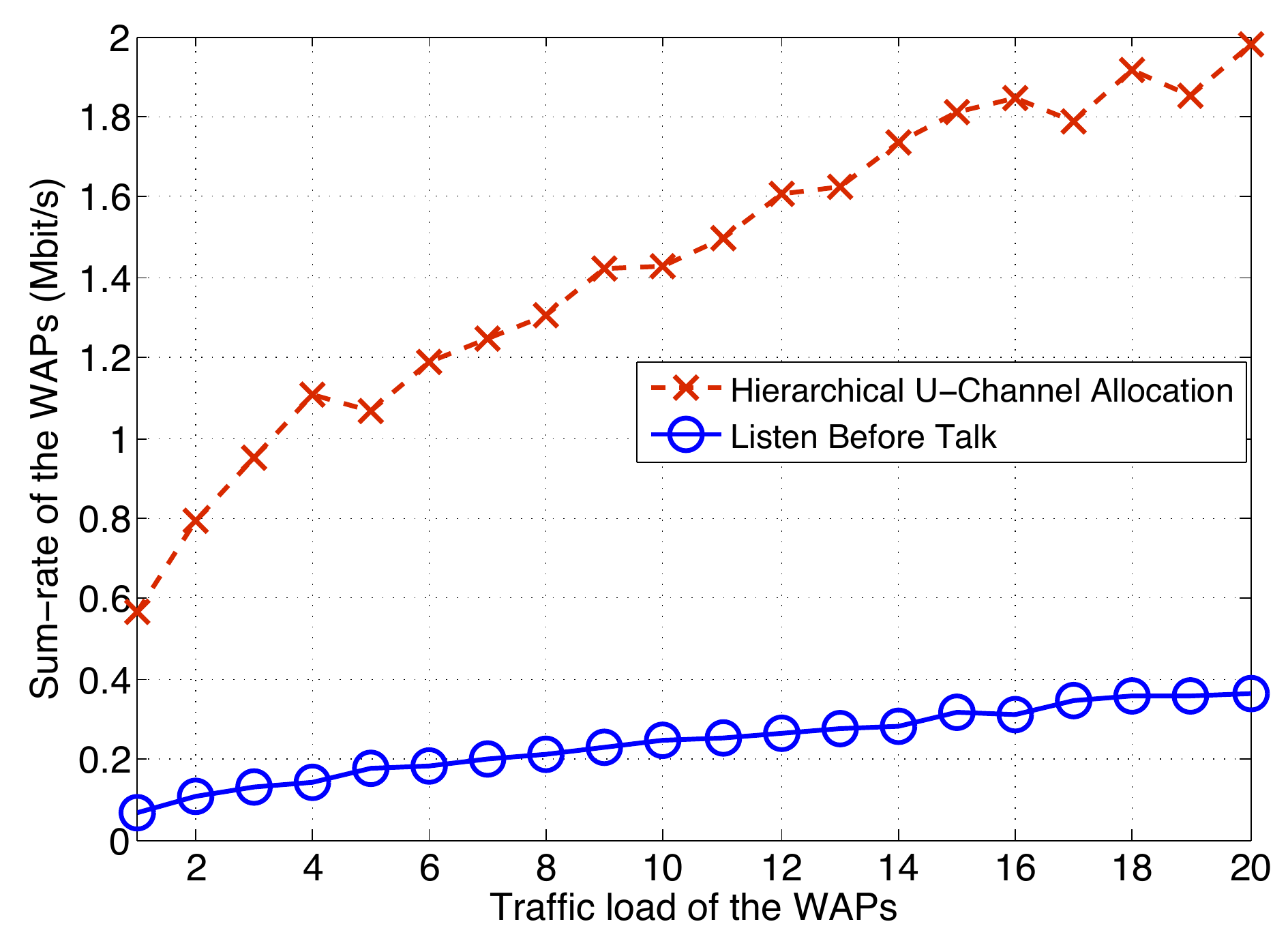}\vspace{-0.1cm}
\caption{Sum-throughput of the WAPs in LBT and the multi-game.}
\label{res3}
\end{figure}
In Figure \ref{res3}, we show the sum-throughput of the WAPs as a function of the WAPs traffic load. We compare the proposed multi-game model with LBT that is used in LAA and LWA. In the case of LBT, the WiFi network saturates faster compared to the multi-game case and this is due to the BSs that try to maximize the amount of offloaded content to the unlicensed bands. Moreover, since all the BSs and WAPs use the same backoff parameters, all the APs have the same priority. In the multi-game mode, the throughput increases by increasing the load in the network until the network saturates. The sum-throughput of the WAPs significantly higher in the multi-game compared to LBT.

\section{Future Outlook}
\label{con}
Clearly, multi-game theory is a promising approach to address the coupled resource allocation problems raised by LTE-U. In this regard, several future opportunities and challenges can be explored. One direction consists in the development of new approaches to analyze the existence, uniqueness, and optimality of the MGE and MGS. Moreover, one must develop practical multi-game learning algorithms, that can properly select, in a fully distributed manner, the most suitable solution for the wireless system, as a whole. On the WiFi side, it is important to ensure through the analysis of multi-game theory that the BSs do not jeopardize the performance of the WiFi users. 

Another important task is to define user-centric approaches that determine which information the terminals should transmit to capture the fact that the dual-mode BSs operate over different bands. Then, the BSs should be able to exploit this information to dynamically optimize the QoS of their LTE users and consequently, the performance of the network. The collected information can also be leveraged to improve the convergence speed of the proposed algorithms.

In this paper, we have proposed a generic game-theoretic framework that can be applied to solve any LTE-U problem. However, the game must be adapted to account for the network characteristics such as the high density of small cell networks and the mobility patterns of the users. To this end, the allocation of the licensed and unlicensed channels can be formulated as mean-field games or on-atomic games. For each scenario, the game must be analyzed and separate analytical results can be found regarding the existence, uniqueness of the MGE and MGS outcomes.\\
\section{ Conclusion}
In this paper, we have proposed a multi-game model to address the new resource allocation problems raised by the introduction of LTE-U into wireless networks. Clearly, multi-game theory can constitute a strong framework for the analysis of LTE-U systems and the development of efficient resource allocation algorithms that account for the inter-network and intra-network interdependences in such systems.

\bibliographystyle{IEEEtran}
\bibliography{references}

\begin{thebibliography}{10}
\providecommand{\url}[1]{#1}
\csname url@samestyle\endcsname
\providecommand{\newblock}{\relax}
\providecommand{\bibinfo}[2]{#2}
\providecommand{\BIBentrySTDinterwordspacing}{\spaceskip=0pt\relax}
\providecommand{\BIBentryALTinterwordstretchfactor}{4}
\providecommand{\BIBentryALTinterwordspacing}{\spaceskip=\fontdimen2\font plus
\BIBentryALTinterwordstretchfactor\fontdimen3\font minus
  \fontdimen4\font\relax}
\providecommand{\BIBforeignlanguage}[2]{{%
\expandafter\ifx\csname l@#1\endcsname\relax
\typeout{** WARNING: IEEEtran.bst: No hyphenation pattern has been}%
\typeout{** loaded for the language `#1'. Using the pattern for}%
\typeout{** the default language instead.}%
\else
\language=\csname l@#1\endcsname
\fi
#2}}
\providecommand{\BIBdecl}{\relax}
\BIBdecl

\bibitem{cisco}
\BIBentryALTinterwordspacing
Cisco, ``Cisco visual networking index: Global mobile data traffic forecast
  update 2014--2019,'' 2015, White Paper. [Online]. Available:
  \url{http://goo.gl/tZ6QMk}
\BIBentrySTDinterwordspacing

\bibitem{qualcomm}
Qualcomm, ``{Qualcomm Research LTE in Unlicensed Spectrum: Harmonious
  Coexistence with Wi-Fi},'' 2014, White Paper.

\bibitem{zhang2015lte}
R.~Zhang, M.~Wang, L.~X. Cai, Z.~Zheng, and X.~Shen, ``{LTE-unlicensed: the
  future of spectrum aggregation for cellular networks},'' \emph{IEEE Wireless
  Communications}, vol.~22, no.~3, pp. 150--159, 2015.

\bibitem{han2012game}
Z.~Han, N.~Dusit, W.~Saad, T.~Ba{\c{s}}ar, and A.~Hj{\o}rungnes, \emph{Game
  theory in wireless and communication networks: theory, models, and
  applications}.\hskip 1em plus 0.5em minus 0.4em\relax Cambridge University
  Press, 2012.

\bibitem{etkin2007spectrum}
R.~Etkin, A.~Parekh, and D.~Tse, ``Spectrum sharing for unlicensed bands,''
  \emph{IEEE Journal on Selected Areas in Communications}, vol.~25, no.~3, pp.
  517--528, 2007.

\bibitem{teng2014sharing}
F.~Teng, D.~Guo, and M.~L. Honig, ``Sharing of unlicensed spectrum by strategic
  operators,'' in \emph{IEEE Global Conference on Signal and Information
  Processing (GlobalSIP)}, 2014, pp. 288--292.

\bibitem{chen2016echo}
M.~Chen, W.~Saad, and C.~Yin, ``Echo state networks for self-organizing
  resource allocation in lte-u with uplink-downlink decoupling,'' \emph{arXiv
  preprint arXiv:1601.06895}, 2016.

\bibitem{bennis2013cellular}
M.~Bennis, M.~Simsek, A.~Czylwik, W.~Saad, S.~Valentin, and M.~Debbah, ``{When
  cellular meets WiFi in wireless small cell networks},'' \emph{IEEE
  Communications Magazine}, vol.~51, no.~6, pp. 44--50, 2013.

\bibitem{Opt16Chen}
M.~Chen, W.~Saad, and C.~Yin, ``Optimized uplink-downlink decoupling in lte-u
  networks: An echo state approach,'' in \emph{IEEE International Conference on
  Communications (ICC), Mobile and Wireless Networks Symposium}, May 2016.

\bibitem{liu2014small}
F.~Liu, E.~Bala, E.~Erkip, M.~C. Beluri, and R.~Yang, ``Small cell traffic
  balancing over licensed and unlicensed bands,'' \emph{IEEE Transactions on
  Vehicular Technology}, vol.~64, no.~12, pp. 5850--5865, 2014.

\bibitem{google}
N.~Jindal and D.~Breslin, ``Lte and wi-fi in unlicensed spectrum: A coexistence
  study,'' Google, White paper, 2015.

\bibitem{zhu2015game}
Q.~Zhu and T.~Ba\c{s}ar, ``Game-theoretic methods for robustness, security, and
  resilience of cyberphysical control systems: games-in-games principle for
  optimal cross-layer resilient control systems,'' \emph{IEEE Control Systems},
  vol.~35, no.~1, pp. 46--65, 2015.

\bibitem{roth2008deferred}
A.~E. Roth, ``Deferred acceptance algorithms: History, theory, practice, and
  open questions,'' \emph{international Journal of game Theory}, vol.~36, no.
  3-4, pp. 537--569, 2008.

\bibitem{saad2009coalitional}
W.~Saad, Z.~Han, M.~Debbah, A.~Hj{\o}rungnes, and T.~Ba{\c{s}}ar, ``Coalitional
  game theory for communication networks,'' \emph{IEEE Signal Processing
  Magazine}, vol.~26, no.~5, pp. 77--97, 2009.

\bibitem{young2004strategic}
H.~P. Young, \emph{Strategic learning and its limits}.\hskip 1em plus 0.5em
  minus 0.4em\relax OUP Oxford, 2004.

\end{thebibliography}
\let\mybibitem\bibitem

\end{document}